\documentclass[review, 3p, number]{elsarticle}
\usepackage{graphicx}
\usepackage{lineno}

\newcommand{\eg}{{\it e.g.}} 
 
\newcommand{\ea}{{\it et al.}}

\newcommand{\vol}[1]{{\bf #1}}

\newcommand{\Kpnn}{$K_L^0 \rightarrow \pi^0 \nu \bar{\nu}$}

\journal{Nucl. Instr. Meth. Phys. Res. A}
\begin{document}
\begin{frontmatter}
\title{Flash ADC data processing with correlation coefficients}

\author{D.~Blyth}
\author{M.~Gibson$^{ }$\fnref{ostate}}
\author{D.~Mcfarland}
\author{J.~R.~Comfort\corref{cor1}}
\cortext[cor1]{Corresponding author.}
\ead{Joseph.Comfort@asu.edu}

\fntext[ostate]{Current address: Physics Dept., Ohio State University, 
Columbus, OH 43210} 
\address{Physics Department, Arizona State University, 
Tempe, AZ 85287-1504, USA}

\begin{abstract}
The large growth of flash ADC techniques for processing signals, 
especially in applications of streaming data, raises issues such 
as data flow through an acquisition system, long-term storage, 
and greater complexity in data analysis.  In addition, experiments 
that push the limits of sensitivity need to distinguish legitimate 
signals from noise. The use of correlation coefficients is examined 
to address these issues.  They are found to be quite successful 
well into the noise region.  The methods can also be extended to 
Field Programmable Gate Array modules for compressing the data flow 
and greatly enhancing the event rate capabilities.
\end{abstract}


\end{frontmatter}


\section{Introduction}
Data acquisition (DAQ) electronics in nuclear and particle physics 
have been shifting from analog to digital methods for many years. 
As part of this process, flash ADC (FADC) chips and Field 
Programmable Gate Array (FPGA) modules are having ever increasing 
use.  Although these chips and modules provide a much greater range 
of data processing options along with the flexibility of firmware 
programming, considerably more data bits often must be passed 
through the electronics hardware, processed in computers, and 
stored for the long term.  Data analysis methods must also be able 
to handle the increased size and complexity of the raw data.  
In some situations, only a small fraction of the total data 
flow may contain meaningful information relevant to the physics 
being pursued.

A common situation is that in which the signal from a detector 
component or a preamplifier is digitized in a FADC.  The signal 
contains information about the energy deposited or collected as 
well as the time.  It may have a shape that depends on the particle, 
or it may represent overlapping pulses.  Although the desired 
information (e.g., energy) may need only a fraction of the data 
bits of the digitized pulse, sophisticated processing is often not 
possible on the fast time scale of the signals.  So the digitized 
pulses are passed on and possibly stored for later analysis.

An additional problem with such signals is that of distinguishing 
a real signal from a noise excursion, perhaps one that is many 
standard deviations from the average.  Precision experiments that 
push to the limits of sensitivity might falsely record such 
excursions while, due to noise fluctuations, miss good signals.

To address such issues, we have been developing methods based on 
correlation coefficient (CC) algorithms.  Such CC methods can be 
used to pick out signals from a stream of input data, to scan 
through atomic or nuclear spectra to identify peaks, to obtain 
initial estimates of the yield and location of a peak for use 
in a detailed fitting program, for pedestal evaluations in the 
presence of noise \cite{noise_pibeta}, among other things.  
Although the context for our work is the KOTO experiment at 
J-PARC \cite{koto_daq}, benefits can also extend to many other 
experiments.

Our initial and primary interest in developing CC methods was for 
use in offline data analysis.  We explore here the ability of CC 
algorithms to identify FADC signals and to distinguish them from 
noise.  Very good estimates of the energy and time of FADC signals 
can be obtained quickly and easily for use in event reconstruction.  
Consideration is also given to ways in which the methods might be 
extended to online applications where only a small fraction of the 
data contains meaningful information, and/or the event rate through 
a DAQ system or the storage of data are limited.

\section{Correlation Functions}
Consider a set of data \{$y_i$\} and a reference peak \{$x_i$\}, 
each with $N$ points.  A simple correlation function is that 
of the cosine similarity function \cite{similarity}
\begin{equation}
\cos\theta = \frac{\sum_i x_i y_i}{\sqrt{\sum_i x_i^2}\:
  \sqrt{\sum_i y_i^2}}
\end{equation}
where each sum is over $N$ points.  This normalized, $N$-dimensional 
scalar product expresses how well two vectors are `aligned' in 
the space. 

More generally, the correlation coefficient $\rho$ is defined as 
\cite{Bevington}
\begin{equation}
\rho = \frac{ N\sum x_i \, y_i - (\sum x_i)\,(\sum y_i)}
{ \left\{ \left[ N\sum y_i^2 - (\sum y_i)^2 \right] \,
\left[ N\sum x_i^2 - (\sum x_i)^2 \right] \right\}^{1/2} } \:.
\end{equation}
Each sum is again over $N$ values.  The value $\rho^2$ is actually 
more meaningful in that it expresses the fraction of the 
variance in the data \{$y_i$\} which is accounted for by the 
hypothesis of the reference peak \{$x_i$\}.  For example, if 
$\rho = 0.7$, about half of the variance in the data would need 
to be ascribed by something other than the reference peak.

A reference peak can be shifted through a data spectrum to identify 
candidate peaks for which $\rho$ at some location exceeds a 
selection criterion.  The sums $\sum x_i$, $\sum x_i^2$, and 
$(\sum x_i)^2$ are all known and fixed.  The sums involving 
$y_i$ alone can be adjusted at each step by dropping off the 
contributions from the first (oldest) data location and adding 
the contributions of the new location.  However, the correlation 
sum $\sum x_i y_i$ must be recomputed at each step.

\section{KOTO Data Acquisition}
To provide a context for the discussion, we consider the KOTO 
experiment which seeks to obtain the first observation of the
\Kpnn\ decay and a measurement of the CP-violating parameter in 
the Standard Model (SM).  
A full description of the KOTO DAQ system, along with an overview 
of the experiment, is provided in Ref.\ \cite{koto_daq}.  Only some 
items relevant to the discussions in this paper are noted here.

The heart of the detector system is a roughly circular array of 2716 
small and large CsI crystals, approximately 90 cm in radius, for 
detecting photons.  The array has a hole of 20 cm by 20 cm, partly 
filled with some veto detectors, for the beams to pass through.  
In front of and surrounding the CsI array are numerous detectors to 
veto events with charged particles and/or photons that miss the array.

\begin{figure}[h]
\centering
\includegraphics*[width=0.40\linewidth]{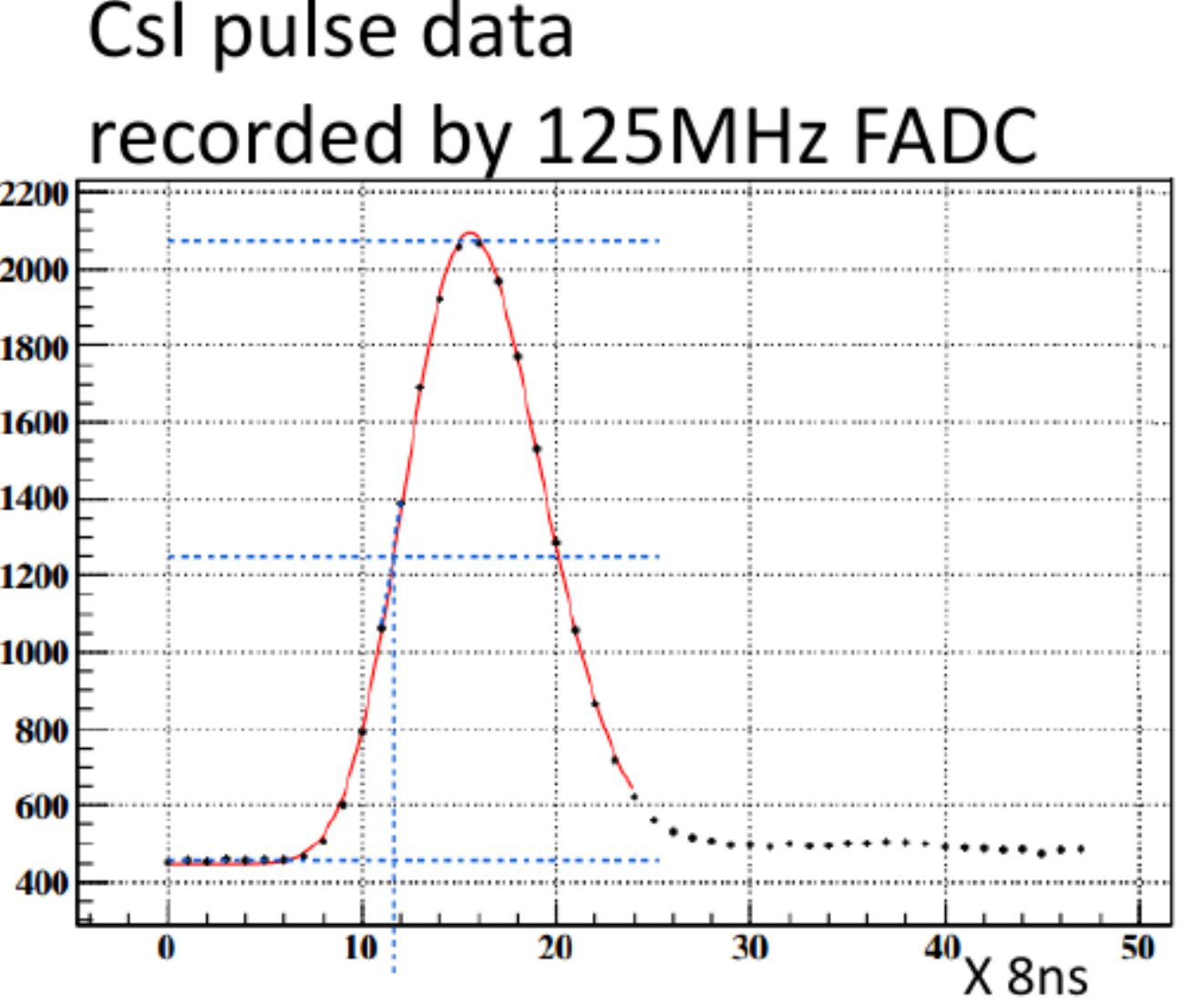}
\caption{An example FADC signal, including an estimated pedestal 
level.}
\label{fig:data_peak}
\end{figure}

Signals from each CsI crystal are sent to custom-made FADC modules, 
each with 16 inputs \cite{Bo07}.  The signals are passed through a 
low-pass filter that is designed to convert the pulse into a 
quasi-Gaussian shape about 45 ns full width at half maximum, and 
with a long and relatively flat tail.  This shape is digitized by 
a 14-bit, 125-MHz chip into 64 time samples (initially 48 samples).  
The top of the peak in each channel is adjusted to be near the 
middle of the time samples, leaving room for samples at each end 
to estimate pedestals.  An example of a FADC signal is shown in 
Fig.\ \ref{fig:data_peak}.

The signals in each module are placed in a 4-$\mu$s pipeline, 
sufficient to retain all data.  Upon receipt of a suitable trigger 
pulse, the FADCs then transmit the pipelined data for all time 
samples of the event to Level 2 (L2) modules.  The L2 modules 
prepare the data for routing through 1-Gb ethernet links to a 
computer array for event building, followed by ethernet transfer 
from J-PARC to the nearby KEK laboratory for long-term storage.  
The DAQ hardware is fully synchronized to an 8-ns clock.

\section{Offline Data Analysis}
For each event, the goal of data analysis is to identify the `hit' 
crystals, their energy deposits, and their times so that clusters 
corresponding to the incident photons can be constructed.  
Kinematic analyses are then applied to reconstruct the $K_L^0$ 
decay modes and their properties.

It is assumed for the discussion here that suitable methods are 
applied to determine pedestals.  Our analyses typically involve 
an algorithm the utilizes the averages of the first 8 and the 
last 8 time samples, with checks on possible accidental peaks 
in those regions \cite{KL016}.  The discussion is also focused 
on the CsI crystals as they determine the kinematics of an event. 

\subsection{Application of the correlation coefficient}
For this work, a model reference peak \{$x_i$\} was constructed 
by averaging the shapes obtained from a large number of peaks 
in the data of an early KOTO test run.  A representation of 
the shape is shown in Fig.\ \ref{fig:data_peak}. 
This parent reference peak consists of 44 samples of real 
(floating point) numbers with a maximum value of 1.0.  
Because the raw data are in the form of 14-bit integers, 
the analysis is made much faster by coding Eq.\ (2) with integer 
arithmetic and bit shifts.  An auxiliary reference peak of 21 
integers is thus used in the data analysis routines.  The data 
for this peak includes all values of the parent reference peak 
down to the 2-3\% level, where the tail becomes flat.  The peak 
is then scaled from the parent reference peak so that the sum 
of the 21 values is $21 \times 512 = 10752$. (See Sec.\ 5.)

To find hits, a pedestal-subtracted time spectrum is prepared 
for each crystal (extended with samples of zero at each end to 
accommodate the samples before the maximum of the reference peak).  
The reference peak is then scanned through a time spectrum, a 
correlation coefficient is calculated at each step, and values 
that exceed the user's criterion are placed in a table along 
with a label of the time sample.  In addition, the CC routine 
automatically calculates the area $A = \sum_i y_i$ of the 
signal peak and the weighted time sum $T = \sum_i i\,y_i$, 
and includes them in the candidate table.  The user's code 
selects the highest CC value from the table, if any, to record 
a hit, along with the area and mean time $T/A$.  The area is 
converted to a value for the peak maximum by a fixed constant. 
The energy is obtained from this peak maximum by applying 
experimental calibration constants.
 
It is to be noted that the information returned to the user is 
obtained directly from the data and is not the result of a 
fitting procedure.  The CC value is simply a measure of the 
quality of the correlation with respect to a reference shape. 
Statistical precision is improved by using a sum over the peak 
distribution rather than a single sample at the maximum.

Our standard analysis code contains some other useful features.  
For example, a time spectrum can be scanned prior to the CC 
analysis to see if it satisfies a tight criterion for having no 
energy, at which point it is removed from further processing.  
A simple method is to find the maximum and minimum values of all 
time samples and test if their difference is less than a specified 
value (\eg, 4-6 $\sigma_{\mathrm{noise}}$) \cite{KL016}.  
Also, because the peaks of all the CsI signals are located near a 
fixed time sample by design, it is not necessary to scan over the 
full time distribution but only over a region around the nominal 
time sample.  This option can eliminate crystals that have only 
out-of-time accidentals.  Two CC criterion levels are available 
for test and development purposes: one in the CC routine to fill 
its table of hit information for the user, and a second, higher 
one for final user selection.  Finally, an energy cut can be 
applied, commonly near the noise floor.

\subsection{Monte Carlo Single Photon Studies}
Real data are often not the best source for testing a correlation 
coefficient method.  The energies and times have uncertainties from 
calibration methods, pedestal values that must be determined 
empirically, non-linearities, noise from known and unknown sources, 
and many other issues.  In short, the true values are unknown and 
the purpose of the data analysis codes is to estimate the best 
values.  The best tests are against Monte Carlo (MC) simulations, 
even with their own uncertainties.

Extensive MC studies were made to test the ability of the CC method 
to properly detect photon events in the CsI array, and especially 
to explore its sensitivity to small signals down to the noise region.  
In these studies, pseudo-FADC data were generated and then processed 
through the offline analysis code in the same way as real data.

In generating the pseudo-FADC data, a pedestal level was generated 
randomly for each crystal.  The values could be selected for the 
analysis, or an algorithm could be used to extract a pedestal from 
the first 8 and/or the last 8 time samples of a time spectrum.  
In the case of the algorithm, the root-mean-square (RMS) deviations 
of pedestals from their averages were about 0.7 ADC counts.

The pedestal fluctuations depend on noise in the system.  From the 
data sheet for the FADC used in the experiment as well as tests 
with the FADC boards \cite{Bo07}, a standard deviation of about 
$\sigma = 2.1$ counts was chosen.  Combined with a model energy 
conversion scale factor of 9 counts/MeV, the noise floor is near 
0.6 MeV (2-3 $\sigma$).  Hence, the MC simulations of energy 
deposits in the crystals were converted to ADC counts, digitized 
with respect to the parent reference peak, and added with Gaussian 
distributed noise fluctuations at each time sample to the pedestal 
value of a crystal.  The signal data were also given random time 
fluctuations within one time sample.  Apart from the effects of 
noise fluctuations, the conversion scale factor cancels in 
round-trip MC simulations.

Single photons for ten energies between 1 and 1000 MeV were 
directed to the CsI crystals.  Except for the lowest energies, 
the photon energy is deposited over several individual crystals.  
The individual deposits and the summed total energy were recorded 
by the MC code.  The analysis code with a CC value $\rho = 0.7$ 
was used to select hits.  A window of about $\pm 4$ time samples 
about the expected location was used, and there was no energy cut.  
The results from the analysis were compared against the original 
MC energy and time values, crystal by crystal for each event, 
to find matched, missed, and false hits.  Full details are given 
in a separate Tech Note \cite{KL018}, with key features 
summarized here.

Even though the CsI crystals have a radiation length of 27$X_0$, 
there is some inefficiency in collecting all of the energy of an 
electromagnetic shower.  Separate MC studies have shown that $e^-$ 
and $e^+$ from the showers (as well as neutrons from photonuclear 
processes) can be emitted and carry energy back upstream.
In addition to the intrinsic loss of energy through physical 
processes, analysis codes may not be able to identify all of the 
deposited energy of a photon due to ambiguities arising from 
photo-statistics, noise, and/or limitations of the detection 
algorithm. At some point, no method that reports energy deposits 
below the noise floor can be considered to be reliable.

With no energy cut on any crystal, the CsI array collected an 
average of 98.0\% to 99.4\% of the incident MC photon energy, 
increasing slowly over the 1--1000 MeV range.  For the same 
range, the analysis code collected an average of 70.2\% to 98.6\% 
of the photon energy over the CsI array.  Note that the 30\% 
discrepancy between the MC and analyzed hits for 1-MeV photons is 
only 0.3 MeV; the discrepancy is 0.8\% (8 MeV) for 1000-MeV photons.  
If crystals with energy deposits less than 1.0 MeV are excluded 
from the sums, both the MC and analysis codes collect nearly equal 
average fractions of the photon energies.  The losses range from 
about 22.5\% or $\sim$0.7 MeV for 3-MeV photons to 2\% or 20 MeV 
for 1000-MeV photons.  Hence, with such a cut, it would be feasible 
on average to correct the energy of a photon cluster obtained from 
data by using an energy-dependent factor determined from MC 
simulations.

On average, the CC method provides very good agreement with the 
energies and times of the hits.  Summing over the energy 
differences, positive and negative, between the individual MC 
deposits and the analyzed hit deposits, the net difference was 
about 0.01 MeV for every photon energy.  The RMS spreads of the 
differences were about 0.13 MeV.  Similarly, the summed time 
differences were about 0.04 time samples with an RMS spread 
typically between 0.60 - 0.70 of the 8-ns time step. 

All of these quantities were obtained with the use of known 
pedestals. If a pedestal algorithm is used, the summed energy 
differences were about 0.07 MeV with a RMS spread of 0.23 MeV, 
while the summed time differences and RMS spreads were unaffected.  
The CC method is not intended to provide high precision for these 
quantities.  Yet it provides excellent estimates for the quantities 
as starting points if needed for detailed fitting routines.

\subsection{Sensitivity in the noise region}
In the previous section, events were considered ``matched" if they 
had the same crystal ID, although the energies were generally 
close as well.  But the energy regions for the missing and false 
hits also need to be examined.

\begin{figure}[h]
\centering
\includegraphics*[viewport=40 40 570 750,angle=90,width=1.0\linewidth]
{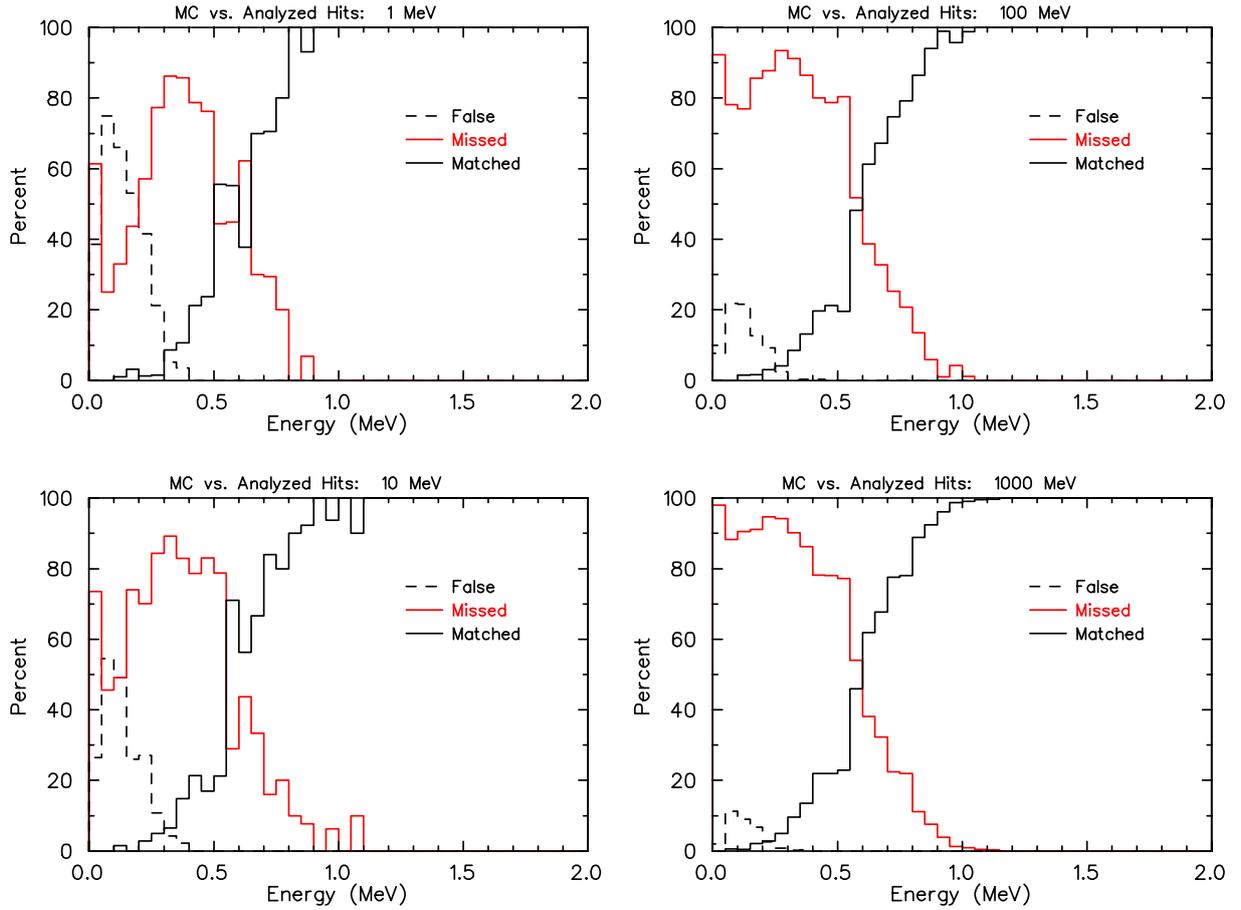}
\caption{Fractions (in percent) vs.\ low-energy deposits within
a cluster of matched missed, and false hits between MC and
analyzed pseudo-data.}
\label{fig:hit_distrib}
\end{figure}

The fraction, in percent, vs.\ low-energy deposits within a 
cluster of matched, missed, and false hits between MC and analyzed 
pseudo-data distributions are shown in Fig.\ \ref{fig:hit_distrib} 
for four of the pure photon energies.  The percent distributions 
are similar in each case, with the lines for the matched and missed 
hits crossing at 50\% near 0.6 MeV (in our model).  This value can 
be interpreted as having an equal probability for either choice: 
below this value, noise prevails; above this value, there is an 
improving likelihood of having a real hit. The missing MC hits 
are nearly all gone by 1 MeV, with a few reaching perhaps 1.5 MeV. 
Hence, one may define the hard noise region $E_{\textrm{hard}} 
< 0.6$ MeV, and a gray region 0.6 MeV $< E_{\textrm{gray}} < 1.0$ 
MeV.  These values agree well with separate studies of the noise 
regions based on pseudo-FADC events with zero energy.

The energy distributions here are based on assumptions in the 
Monte Carlo simulations of the overall energy calibration of 
FADC counts per MeV.  They may be different for KOTO or any 
other experiment.  In addition, photo-statistics can affect 
the behavior at these low energies, and are not included in 
the model.  Such fluctuations would increase the energy values 
for the hard and gray regions. 

\section{Online Data Compression}
All of our MC calculations showed that very few CsI crystals have 
energy deposits in an event.  Out of 2716 crystals, a 1-MeV photon 
deposits energy on average in 1.4 crystals, while a 1000-MeV photon 
would have hits in 48 crystals.  Even complicated multi-photon 
decays had deposits only up to 200 hits per event, with average 
values of about 100-150 hits per event.  Hence, 95\% or more of the 
event data have no direct physics content.  If crystals with real 
energy deposits can be cleanly distinguished from those with only 
noise, substantial enhancements in DAQ event rates, along with 
savings in analysis time and data storage, can be implemented.
Large compression factors of 20, 30, or more might be possible.

To accomplish such a task, the correlation coefficient would need 
to  be recomputed every time sample (8 ns for KOTO) on data in a 
FADC pipeline.  Although challenging, a rearrangement of Eq.\ (2) 
along with the use of integer arithmetic and bit shifts has 
provided a solution that can be encoded in the FADC firmware.

It is not necessary to know the actual CC value $\rho$ to make a 
decision about whether to retain some data or not, but only that it 
exceeds a preset value.  In that case, it will be convenient to 
square Eq.\ (2) and rearrange it in a slightly different form.  
To ensure a positive correlation, require that the numerator of 
Eq.\ (2) before squaring be greater than zero.  Then, defining 
$X_i = N x_i$, 
\begin{equation}
\left[\, \sum X_i \, y_i - \frac{1}{N}(\sum X_i) (\sum y_i) \, 
\right]^2 = \left\{ \rho^2 \left[ \frac{1}{N}\sum X_i^2 - 
\frac{1}{N^2}(\sum X_i)^2 \right] \right\} 
\left[ N\sum y_i^2 - (\sum y_i)^2 \right] \:.
\end{equation}
As mentioned in Sec.\ 4.1, $\sum x_i$ in our model is normalized 
to 512 ($2^9$), so $\sum X_i = 10,752$.  This scaling is needed to 
obtain sufficient accuracy with integer arithmetic for small signals. 
The factor in curly braces can be precomputed once.  The selection 
criterion is met if the left-hand side is greater than the right-hand 
side.  It is not necessary to complete the calculations over the 
full dynamic range of the signals because those that exceed some 
level will certainly be passed.  If the test fails, a compression 
bit for the corresponding ADC input to the module can be set, and 
referenced by subsequent data stages to recognize missing data.

The Monte Carlo studies have shown that a value of $\rho = 0.7$ 
is generally successful.  It has the ability to reach well into 
the noise region without picking up an excessive number of false 
hits.  A more conservative value of 0.6 may be better for FADCs, 
to preserve marginal cases for further analysis. For the MC data, 
the extra hits with $\rho = 0.6$ all had energies below 1 MeV.  
Values below 0.6 tended to be excessively sensitive to noise.  
As with the offline case, a time window can be used for the CC 
scans to suppress accidentals.

Because the correlation coefficient function is complex and can 
take up considerable space in a FPGA, replacing it with the 
similarity function Eq.\ (1) may be satisfactory.  In offline 
comparisons, very little difference has been found between the 
results of the two functions.

\section{Summary and Conclusions}
Many experiments need to be sensitive to signals that border on or 
are embedded in the noise region.  A common approach is to test for 
an excursion that is a few noise standard deviations above the 
pedestal level to define a signal.  This method, however, can be 
compromised by additional random fluctuations in the data, such as 
accidentals, or the pedestal level.  Also, excursions a few time 
samples from the correct time can occur.

If the signals have an approximately fixed shape apart from magnitude 
(or can be transformed to such a shape), a correlation comparison 
against a reference peak is a far better way to identify a real 
signal.  To dig a signal pattern out of noise fluctuations with 
some level of confidence, there is little choice but to use such 
comparisons.  At the lower values of $\rho$, the CC method is 
effectively a `bump' detector for low signals.

Detailed studies with simulated data have established that the 
correlation-coefficient (or similarity) method is very successful 
in identifying energy deposits (hits) in detector elements well into 
the noise region.  It provides great benefits especially for 
offline analysis.  It easily provides very good estimates of the 
signal energy and time that are needed for subsequent detailed 
fitting, and optionally for reducing the amount of stored data.

For online use in FADC modules, the application of the method should 
be sufficiently conservative to preserve possible hits of small 
signals for later analysis, but also to provide a high level of 
data compression.  As with other possible methods of working in 
the noise region, there is some risk of losing a real signal.  
The use of the method must be carefully tailored to the design 
requirements of the experiment, especially the balance between 
the required signal range, photo-statistics, and other sources 
of noise.

\section*{Acknowledgment}
We thank our colleagues on the KOTO experiment for many thoughtful
questions and comments during the development.  This work was supported 
in part by the DOE award DE-SC0002644 through a subcontract from the 
University of Michigan and DOE award DE-SC0006497 to Arizona State 
University.


\end{document}